\documentclass[11pt]{article}

\usepackage[left=2cm, right=2cm, top=2.5cm, bottom=2.5cm]{geometry}
\usepackage[x11names]{xcolor}
\usepackage{fancyhdr, amssymb, cancel, amsmath, graphicx, pgfplots, tikz}

\newcommand{\stylecolor}{black}

\usepackage[colorlinks=false, hyperindex=true, linktocpage=true]{hyperref}

\usepackage[explicit]{titlesec}

\newcommand*\sectionlabel{}
\titleformat{\section}
  {\gdef\sectionlabel{}
   \Large\bfseries\scshape}
  {\gdef\sectionlabel{\thesection. }}{0pt}
  {\begin{tikzpicture}[remember picture,overlay]
	\draw (-0.2, 0) node[right] {\textsf{\sectionlabel#1}};
	\draw[thick] (0, -0.4) -- (\textwidth, -0.4);
       \end{tikzpicture}
  }
\titlespacing*{\section}{0pt}{15pt}{20pt}

\newcommand*\subsectionlabel{}
\titleformat{\subsection}
  {\gdef\subsectionlabel{}
   \large\bfseries\scshape}
  {\gdef\subsectionlabel{\thesubsection.\ \  }}{0pt}
  {\begin{tikzpicture}[remember picture,overlay]
    	\draw (-0.15, 0) node[right] {\textsf{\subsectionlabel#1}};
       \end{tikzpicture}
  }
\titlespacing*{\subsection}{0pt}{10pt}{10pt}

\pgfplotsset{every axis legend/.append style={at={(1.02,1)},anchor=north west}}

\newcommand{\titletext}{Exact mean field dynamics	 for epidemic-like processes on heterogeneous networks}

\begin{document}
\thispagestyle{empty}

\begin{equation*}
\begin{tikzpicture}
\draw (0.5\textwidth, -3) node[text width = \textwidth] {{\huge \begin{center} \color{\stylecolor} \textsf{\textbf{\titletext}} \end{center}}}; 
\end{tikzpicture}
\end{equation*}
\begin{equation*}
\begin{tikzpicture}
\draw (0.5\textwidth, 0.1) node[text width=\textwidth] {\large \color{black} \textsf{Andrew Lucas}};
\draw (0.5\textwidth, -0.5) node[text width=\textwidth] {\small \textsf{Jefferson Physical Laboratory, Harvard University}};
\draw (0.5\textwidth, -1) node[text width=\textwidth] {\small \textsf{Department of Physics, Stanford University}};
\end{tikzpicture}
\end{equation*}
\begin{equation*}
\begin{tikzpicture}
\draw (0.5\textwidth, -6) node[below, text width=0.8\textwidth] {\small  We show that the mean field equations for the SIR epidemic can be exactly solved for a network with arbitrary degree distribution.  Our exact solution consists of reducing the dynamics to a lone first order differential equation, which has a solution in terms of an integral over functions dependent on the degree distribution of the network, and reconstructing all mean field functions of interest from this integral.  Irreversibility of the SIR epidemic is crucial for the solution.  We also find exact solutions to the sexually transmitted disease SI epidemic on bipartite graphs, to a simplified rumor spreading model, and to a new model for recommendation spreading, via similar techniques.  Numerical simulations of these processes on scale free networks demonstrate the qualitative validity of mean field theory in most regimes.};  
\end{tikzpicture}
\end{equation*}
\begin{equation*}
\begin{tikzpicture}
\draw (0, -13.1) node[right] {\textsf{Correspond with: }\texttt{lucas@fas.harvard.edu}};
\draw (\textwidth, -13.1) node[left] {\textsf{\today}};
\end{tikzpicture}
\end{equation*}

\tableofcontents

\pagestyle{fancy}
\fancyhead{}

\fancyhead[L] {\textsf{\titletext}}
\fancyhead[R] {\textsf{\thepage}}
\fancyfoot{}

\section{Introduction}
Over the last decade, there has been increasing interest in how network heterogeneity may affect nonequilibrium dynamics in qualitative ways \cite{barratbook}.   One of the simplest and most important examples has been the susceptible to infected (SI) and suspectible to infected to removed (SIR) epidemic models, famous from epidemiology \cite{murray}, which model disease outbreaks in populations.   A decade ago,  \cite{pastor} first demonstrated that heterogeneous networks can fundamentally alter the dynamics of these processes in qualitative ways --  in particular, the epidemic threshold vanishes on scale free graphs of degree $\gamma \le 3$,  so epidemics always infect a nontrivial fraction of nodes on an infinite graph, with finite size corrections later shown to be extremely small \cite{pastor10}.\footnote{In this paper, we will often casually say 	``no epidemic threshold" when we are really referring to epidemic thresholds which vanish rapidly with $N$, the size of the network.}  Later, in \cite{pastor2} it was shown that the removal of this threshold also corresponds to faster than linear epidemic growth on such graphs.   Many authors have all explored various aspects of the dynamics of epidemic spreading.   \cite{boguna, pastor3, pastor2, morenoy, gomez} extend analysis of mean field theory, and within this framework \cite{may, newman} discuss the late time behavior of epidemics on scale free graphs, with \cite{marder, noel} introducing some dynamical aspects.   \cite{volz, millernote}, closest in spirit to this work, present reductions of the dynamical equations, although their approach is quite different.   Mathematicians have used many complicated techniques to obtain information about generalizations of such solutions to more complicated epidemic types \cite{valls, nucci}, but have typically avoided studying the complication of adding an entire network structure.  \cite{hetero} introduces an extension where bipartite graph structure can be reasonably accounted for by mean field theory, leading to a model of sexually transmitted disease (STD) epidemics.

In addition to processes which may be well modeled by the SIR epidemic, there are many others which share the same structure of the SIR epidemic -- irreversible flow from $\mathrm{S}\rightarrow\mathrm{I}\rightarrow\mathrm{R}$.    A related example of such a process is that of rumor spreading \cite{moreno1, moreno2}, which is similar to that of SIR epidemic spreading but with a ``death rate" which is proportional to the current number of infected edges.   A slightly more complicated version of the model also allows for the infected nodes to die on their own \cite{nekovee}, but the fundamental difference between this model and the epidemic is captured without this term.     Other irreversible processes, such as a new model for recommendation spreading in a population \cite{blattner}, are also very similar, even if they do not have an identical $\mathrm{S}\rightarrow\mathrm{I}\rightarrow\mathrm{R}$ structure.

In this paper, we will present mean field dynamical solutions to the following 4 models:  the SIR epidemic, the SI epidemic on bipartite graphs,  a simplified model of rumor spreading in which only infected edges can induce transitions to the removed state, and the  recommendation spreading model.   These solutions should, for all intents and purposes, be regarded as exact -- the only approximation that they require is mean field theory, and they allow for reconstruction of all dynamical quantities of interest within the scope of mean field theory (most easily by numerical methods).   For each model, the exact solution can be found for arbitrary degree distribution, when written in the form of an integral over a function defined based on the degree distribution of the underlying network.   We will typically make some simplifying approximations to reduce the amount of work we have to do in analyzing the theoretical dynamics, but we stress that these approximations can be removed.

There are numerous reasons why the existence of such exact mean field solutions for arbitrary (mean field) networks is helpful.   Other exact solutions have typically either focused only on the behavior at very late times \cite{newman}, or focused on very special types of graphs like the nearest neighbor 1D lattice \cite{williams}, or expressed as series solutions, which obscure the physical meaning of the solution \cite{khan}. Most importantly, the exact solution allows one to determine the accuracy of mean field theory, beyond a comparison of scaling behaviors.   Furthermore, an exact solution provides dynamical information about the nature of the epidemic away from the fixed points of the dynamics, as well as precise information about the dynamics in regimes where linearized approximations break down, and we will indeed find  more precise answers than we have found in the literature.   We will present a basic analysis of the resulting equations as well as compare our results to numerical simulations, which are typically quite accurate.   For simplicity, we will almost always work with scale free graphs, where the exact solution can be expressed in terms of integrals over incomplete $\Gamma$ functions with well understood properties -- furthermore, such graphs capture the essence of how network structure can dramatically change the qualitative dynamics.

The paper is organized as follows.  Section \ref{epsec} discusses the epidemic models, while Section \ref{rusec} describes the rumor spreading models and Section \ref{sad} discusses the recommendation model;  Section \ref{concsec} presents a discussion of the work.   Numerical results are presented as we discuss the theory.

As this work was being finalized, we discovered a recent series of papers \cite{jc1, jc2, jc3} which discuss modeling variations of the SIR epidemic by reduction of the dynamics to finite sets of ODEs, using a technique somewhat related to ours.  The focus of this work is quite different, emphasizing scaling behavior and asymptotic dynamics, as well applying this technique to models beyond the scope of epidemic spreading.

\section{SIR Epidemics}\label{epsec}
We begin by discussing the exact mean field solutions, and numerical corroborations of these solutions, for the epidemic spreading models.  We first discuss the general structure of an ``epidemic like" process, then move on to the SIR epidemic, then describe why the irreversibility is so crucial, and finally discuss the SI STD epidemic.
\subsection{General Overview of Epidemic Processes on Networks}
This section is meant as a brief review of the nature of an epidemic-like process on a network, and the well-versed reader may happily skip it or  skim it to ensure that he understands our notation.

We begin by quickly reviewing what we mean by a network, or graph.  An (undirected) graph is a set of vertices $V$, along with a set of edges $E$, with an edge $e\in E$ associated to a pair of vertices:  $e=(uv)=(vu)$ with $u,v\in V$.    The degree of a vertex (or node) $v$, which we will label $k_v$, is the number of edges in $E$ with one of the ends of the edge being $v$.   
The SIR epidemic is a stochastic process defined on such a network.  The state space for this stochastic process is given by $\lbrace \mathrm{S}, \mathrm{I}, \mathrm{R}\rbrace^{|V|}$ -- i.e., each node can exist in state S, I or R.   In theory, the SIR epidemic is a continuous stochastic process, with the rate of transition between states being defined as follows:  if two graph configurations differ by more than 1 node, then no transitions are allowed.   If the graphs differ by one node, than the following transitions are allowed: \begin{equation}
\text{for each node } v\in V: \;\;\;\;\; \left\lbrace\begin{array}{l} v: \mathrm{S}\rightarrow\mathrm{I} \text{ with rate } k_v\theta_v \\ v: \mathrm{I}\rightarrow\mathrm{R} \text{ with rate } \lambda   \end{array}\right.,
\end{equation}where \begin{equation}
\theta_v \equiv \frac{\text{number of edges which point from }v \text{ to a node in state I}}{k_v}
\end{equation}Note that we have chosen to measure time in units where the rate of transition from S to I is 1, per edge.

The intuition for the above process is straightforward.  If a node is an S, it is susceptible to becoming infected, which occurs by an interaction with an infected neighbor.  The more infected neighbors the node has, the more likely the node is to catch the infection from one of them -- we assume this rate is linear.  We then assume that a node dies with a constant rate once they catch the disease.   There are many obvious variations on such a process, although most of them will not be likely to have an exact solution of the type found in this paper.   We will consider a few simple processes of this form which do have such exact solutions.

It is well-known that mean field theory is typically a far better approximation to dynamical processes on such networks than on a graph like a hypercubic lattice, as the random structure of the graph, and the large number of edges, mean that the network itself helps to ``average" over states \cite{barratbook}.   In this paper, we will always assume that $|V|\rightarrow\infty$ (the number of nodes is getting infinitely large) -- this is the regime where mean field theory should work best.   Mean field theory will treat all nodes with the same $k_v$ as being the same, and so all we will care about   is $\rho_k$, the fraction of nodes in $V$ which have $k_v=k$, and $S_k$, $I_k$ and $R_k$, the fraction of nodes which have $k$ edges which are in state S, I, or R respectively.   Conservation of probability tells us that \begin{equation}
S_k + I_k+R_k=1
\end{equation}  and so we can neglect the dynamics of $R_k$.  The other key approximation of mean field theory will be that\begin{equation}
\theta_v = \theta \equiv\left[\sum k\rho_k\right]^{-1} \sum k\rho_k I_k \equiv \frac{\langle kI_k\rangle}{\langle k\rangle},   \label{ourtheta}
\end{equation}where we are using angle brackets to denote averages with respect to the distribution $\rho_k$.\footnote{(\ref{ourtheta}) is a bit simplistic, because since every infected node (other than a starting ``seed" infected node) was infected by contact with some other infected node, in reality an infected node with $k$ edges could at most transmit the infection to $k-1$ other states.   However, we will only simulate things on graphs where each node has at least 5 or so edges, and this will not turn out to have a very large qualitative, or quantitative, impact on the discussion.  It is also be very straightforward to remove this approximation, at the expense of introducing some more terms into the equations.}

\subsection{Solution for Scale Free Graphs}
\begin{figure}[here]
\centering
\begin{tikzpicture}

\begin{axis}[width=7cm, height=6cm, xlabel=$k$, ylabel=$-\log S_k\;\;$, ylabel style=sloped like x axis]

\pgfplotstableread{nu35n5000z.txt}\datatable
\addplot[color=violet, mark=*, only marks] table[x index=0, y index=9] from \datatable;
\addplot[color=blue, mark=*, only marks] table[x index=0, y index=12] from \datatable;
\addplot[color=red, mark=*, only marks] table[x index=0, y index=14] from \datatable;

\addplot[color=red, domain=0:40] {0.0194*x};
\addplot[color=blue, domain=0:40] {0.0117*x};
\addplot[color=violet, domain=0:40] {0.005*x};
\end{axis}

\end{tikzpicture}
\caption{$-\log S_k$ as a function of $k$ at various times.   We generated scale free graphs with $N=5000$ nodes, degree $\gamma=3.5$, and death rate $\lambda=9$, and averaged over 200 trials.}
\label{zexpsir}
\end{figure}
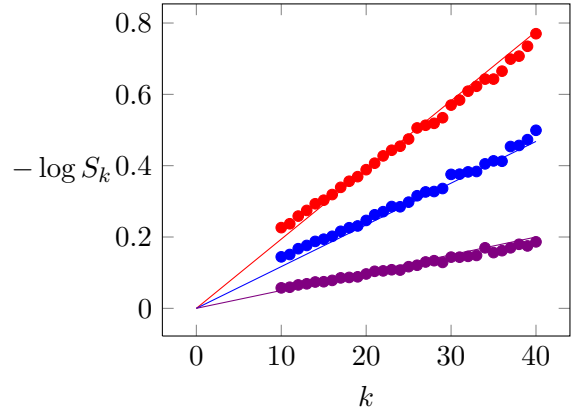

The mean field equations of the SIR epidemic are easy to write down, given the rules above: \begin{subequations}\begin{align}
\dot{S}_k &= -k\theta S_k,  \label{sk} \\
\dot{I}_k &= k\theta S_k - \lambda I_k \label{ik}.
\end{align}\end{subequations}Now, let us reduce this infinite set of dynamical equations, assuming that all nodes in the graph have at least $m$ edges.   We begin with (\ref{sk}): \begin{equation}
\frac{\dot{S}_k}{\dot{S}_m} = \frac{\mathrm{d}S_k}{\mathrm{d}S_m} = \frac{-k\theta S_k}{-m\theta S_m} = \frac{k}{m}\frac{S_k}{S_m}.
\end{equation}  This can be easily integrated to give, if we assume that $S_k(0)\approx S_m(0)\approx 1$: \begin{equation}
S_k(t) = S_m(t)^{k/m}. \label{skt}
\end{equation}
For later convenience, we will introduce the variable \begin{equation}
z(t) = -\log S_m(t),  \label{zsm}
\end{equation}and we find we have reduced (\ref{sk}) to \begin{equation}
\dot{z} = m \theta. \label{dz}
\end{equation}
As we show in Figure \ref{zexpsir}, numerical simulations suggest that (\ref{skt}) becomes very quickly quantitatively true for a decent range of $k$ as soon as the epidemic takes off.   We use scale free graphs, with \begin{equation}
\rho_k \sim \Theta(k-m) k^{-\gamma}  \label{rhosf}
\end{equation} for simulations for the entirety of this paper, as that is where the dynamics becomes most interesting, and where our mean field solutions will become easier to write down.   In all of our simulations, we use $m=10$.\footnote{We checked that this assumption did not lead to any qualitative changes in behavior -- e.g., if $m=5$ or $m=20$, the dynamics are very similar.}  To generate quality scale free graphs, we use the preferential attachment algorithms of \cite{krapivsky}.\footnote{Other papers, e.g. \cite{pastor2}, show that the specific algorithm used to generate a scale free graph does not result in any qualitative change to the dynamics, so we will not worry about this point.}  For a bit larger $m$, the values of $z$ become significantly higher, but this is a numerical fragment ($-\log 0 = \infty$ -- i.e., all nodes of a given connectivity have been infected or removed), and so we have truncated these unphysical values from our graph.

Now, we turn to (\ref{ik}), and we find an equation for $\dot{\theta}$: \begin{equation}
\dot{\theta} = \sum \frac{k\rho_k \dot{I}_k}{\langle k\rangle} = \sum \frac{\rho_k}{\langle k\rangle} \left[k^2\theta S_k - \lambda k I_k\right] =\left[\sum \frac{k^2}{\langle k\rangle} \rho_k S_k - \lambda \right]\theta.  \label{dtheta}
\end{equation}Now, using that $S_k = \mathrm{e}^{-kz/m}$:  we find that:\begin{equation}
\frac{\dot{\theta}}{\dot{z}} = \frac{\mathrm{d}\theta}{\mathrm{d}z} = \frac{1}{m\langle k\rangle} \sum k^2\rho_k \mathrm{e}^{-kz/m} -\frac{\lambda}{m},
\end{equation}which implies that\begin{equation}
\theta(z) = \sum \frac{k\rho_k}{\langle k\rangle} \left(1-\mathrm{e}^{-kz/m}\right)-\frac{\lambda}{m}z = 1-\frac{\lambda z}{m} - \sum \frac{k\rho_k}{\langle k\rangle} \mathrm{e}^{-kz/m}
\end{equation}

Now, using (\ref{rhosf}), let us approximate that our graph is scale free.   This will turn out to make $\theta(z)$ have (approximately) an exact expression in terms of well-understood functions:
\begin{align}
\theta +\frac{\lambda z}{m}-1 &\approx - \int\limits_m^\infty \mathrm{d}k \frac{(\gamma-1)m^{\gamma-1}}{k^\gamma} \left[\frac{\gamma-1}{\gamma-2}m\right]^{-1} k\mathrm{e}^{-kz/m} \notag \\
&=- (\gamma-2)  \int\limits_z^\infty \mathrm{d}x \;  \mathrm{e}^{-x} \frac{1}{z}\left(\frac{z}{x}\right)^{\gamma-1} = -(\gamma-2) z^{\gamma-2}\Gamma(2-\gamma ,z)
\end{align}Note that the $\gamma$ and $m$ dependent factors we have introduced are so that the probability distributions integrate to 1.  We have also used identities out of \cite{abramowitz}: here $\Gamma(a,z)$ is the upper incomplete $\Gamma$ function.   Using another identity we find \begin{equation}
\theta(z) = z^{\gamma-2} \Gamma(3-\gamma,z) + 1 - \mathrm{e}^{-z} - \frac{\lambda}{m}z.
\end{equation}We then find that we have reduced the dynamics, under fairly benign approximations, to a very simple form: \begin{equation}
\dot{z} = mz^{\gamma-2}\Gamma(3-\gamma,z) + m\left(1-\mathrm{e}^{-z}\right) - \lambda z.  \label{dzdt}
\end{equation}
We can thus write down the exact mean field solution, (within our mild approximations): \begin{equation}
t = \int\limits_{z(0)}^z \frac{\mathrm{d}z^\prime}{m(1-\mathrm{e}^{-z^\prime} + z^{\prime(\gamma-2)}\Gamma(3-\gamma,z^\prime)) - \lambda z^\prime}  \label{exact1}
\end{equation}Note that we require a very small $z(0)$ factor to regularize divergences -- we will discuss the physical consequences of this shortly.  The physical meaning of this factor, as the initial condition of the dynamics, is clear.   We should also note that by simply replacing the denominator of (\ref{exact1}) with $m \theta(z)$, we have the exact solution for an arbitrary graph.

While we have an exact solution, since it involves an integral, it is easier to just analyze (\ref{dzdt}).   It is straightforward to justify by considering the asymptotic behaviors of the various terms that there are at most two fixed points:  $z=0$ is always a fixed point, and if it is unstable, there is an absolutely stable fixed point at $z=z^*>0$ some finite point.   To analyze the stability of the $z=0$ fixed point, about which dynamics occur, we re-write (\ref{dzdt}) as $z\rightarrow 0$ \begin{equation}
\dot{z} \approx z\left[m\frac{\Gamma(3-\gamma,z)}{z^{3-\gamma}} + m-\lambda\right].
\end{equation}
Suppose that $\gamma>3$.  Using yet another identity from \cite{abramowitz} concerning the small $z$ behavior of the $\Gamma$ function term, we find that \begin{equation}
\dot{z} \approx z \left[\frac{\gamma-2}{\gamma-3}m  - \lambda \right]
\end{equation}which implies the existence of an epidemic threshold: \begin{equation}
\lambda_{\mathrm{c}} = \frac{\gamma-2}{\gamma-3}m.
\end{equation}For $\lambda<\lambda_{\mathrm{c}}$, epidemics will not spread, whereas they will for $\lambda>\lambda_{\mathrm{c}}$.   Since the fixed point at finite positive  $z$ is always absolutely stable, we conclude that for $\lambda\ne\lambda_{\mathrm{c}}$, the dynamics are always linear near fixed points.  Since these are the slow points of the dynamics, we conclude that the time scales of the dynamics, the spreading time $\tau_{\mathrm{spread}}$, and the ending time $\tau_{\mathrm{end}}$, should be \begin{equation}
\tau_{\mathrm{spread}} \sim \tau_{\mathrm{end}} \sim \int\limits_{1/N}^{\mathrm{O}(1)}\frac{\mathrm{d}z}{z} \sim \log N.
\end{equation}
Of course, we do not take our precise approximation of $\lambda_{\mathrm{c}}$ too seriously, but the key point is simply that there is an epidemic threshold, and a finite time scale of the epidemic dynamics, when $\gamma>3$.  This fact is well known \cite{pastor2}.

Now, let us consider the case where $\gamma< 3$.  Here, the $\Gamma$ function ratio is now divergent as $z\rightarrow 0$, and so the dominant term of the dynamics is \begin{equation}
\dot{z} \sim z^{\gamma-2}.
\end{equation}From this we find the spreading time scale is \begin{equation}
\tau_{\mathrm{spread}} \approx \int\limits_{1/N}^{\mathrm{O}(1)} \frac{\mathrm{d}z}{z^{\gamma-2}} \sim \left. z^{3-\gamma}\right|_{1/N}^{\mathrm{O}(1)} = \mathrm{O}(1).
\end{equation}

In the case  of $\gamma=3$, we have that $\Gamma(0,z)\sim -\log z$, and so denoting \begin{equation}
y\equiv -\log z,
\end{equation}we find that we can approximate the dynamical equation by \begin{equation}
\dot{y} \approx -y
\end{equation}for large $y$, with initial condition $y_0\sim \log N$.   This immediately gives us that \begin{equation}
\tau_{\mathrm{spread}} \sim \log\log N.
\end{equation}

It was argued heuristically, and shown numerically, in \cite{pastor2} that the growth of epidemics was faster than linear for scale free graphs with $\gamma\le 3$.  Here, however, we have a more precise claim that the time scale of epidemic spreading is in fact independent of the size of the network (except in the special case $\gamma=3$).   We similarly find for this case that $\tau_{\mathrm{end}} \sim \log N$.

Now that we have an exact solution and understand its important properties, the most important question is whether or not we can use the exact solution to actually determine the dynamics of various functions of interest: $S_k(t)$, $I_k(t)$ and $R_k(t)$.  Of course, it will suffice to find the first two, and the first follows directly from (\ref{zsm}) and (\ref{skt}).   To find $I_k(t)$, we can use the following trick: \begin{equation}
\frac{\mathrm{d}}{\mathrm{d}t} \left(\mathrm{e}^{\lambda t} I_k(t)\right) = k\mathrm{e}^{-kz(t)/m} \theta(z(t)).
\end{equation}Having found $z(t)$, we can recover \begin{equation}
I_k(t) = \int\limits_0^t \mathrm{d}s\; \mathrm{e}^{-\lambda (t-s)} k\mathrm{e}^{-kz(s)/m} \theta(z(s)).
\end{equation}where we have approximated that $I_k(0)\approx 0$.  It is likely not possible to do these integrals by hand, but they could be done numerically.

\begin{figure}[here]
\centering
\begin{tikzpicture}
\begin{axis}[width=7cm, height=5.5cm, xlabel=$t$, ylabel=$z$, ylabel style=sloped like x axis]

\pgfplotstableread{nu35l2.txt}\datatable
\addplot[color=violet, very thick] table[x index=0, y index=1] from \datatable;

\pgfplotstableread{nu35l4.txt}\datatable
\addplot[color=blue, very thick] table[x index=0, y index=1] from \datatable;

\pgfplotstableread{nu35l8.txt}\datatable
\addplot[color=red, very thick] table[x index=0, y index=1] from \datatable;

\pgfplotstableread{testep352.txt}\datatable
\addplot[color=violet!60!white, very thick, dotted] table[x index=0, y index=1] from \datatable;

\pgfplotstableread{testep354.txt}\datatable
\addplot[color=blue!60!white, very thick, dotted] table[x index=0, y index=1] from \datatable;

\pgfplotstableread{testep358.txt}\datatable
\addplot[color=red!60!white, very thick, dotted] table[x index=0, y index=1] from \datatable;

\draw (axis cs: 1.5, 0.3) node[left] {$\gamma=3.5$};
\end{axis}
\begin{scope}[xshift=6.2cm]
\begin{axis}[width=7cm, height=5.5cm, xlabel=$t$]

\pgfplotstableread{nu25l2.txt}\datatable
\addplot[color=violet, very thick] table[x index=0, y index=1] from \datatable;
\addlegendentry{$\lambda=2$};

\pgfplotstableread{nu25l4.txt}\datatable
\addplot[color=blue, very thick] table[x index=0, y index=1] from \datatable;
\addlegendentry{$\lambda=4$};

\pgfplotstableread{nu25l8.txt}\datatable
\addplot[color=red, very thick] table[x index=0, y index=1] from \datatable;
\addlegendentry{$\lambda=8$};

\pgfplotstableread{testep252.txt}\datatable
\addplot[color=violet!60!white, very thick, dotted] table[x index=0, y index=1] from \datatable;

\pgfplotstableread{testep254.txt}\datatable
\addplot[color=blue!60!white, very thick, dotted] table[x index=0, y index=1] from \datatable;

\pgfplotstableread{testep258.txt}\datatable
\addplot[color=red!60!white, very thick, dotted] table[x index=0, y index=1] from \datatable;

\end{axis}
\end{scope}
\end{tikzpicture}
\caption{Comparison of mean field theory prediction for $z(t)$ to numerical simulations.  We used $N=2000$, $m=10$ and averaged over 50 trials.   The unphysical jumps in $\lambda=2$ dynamics are due to trials where all nodes with $k=10$ became infected.}
\label{fig1}
\end{figure}
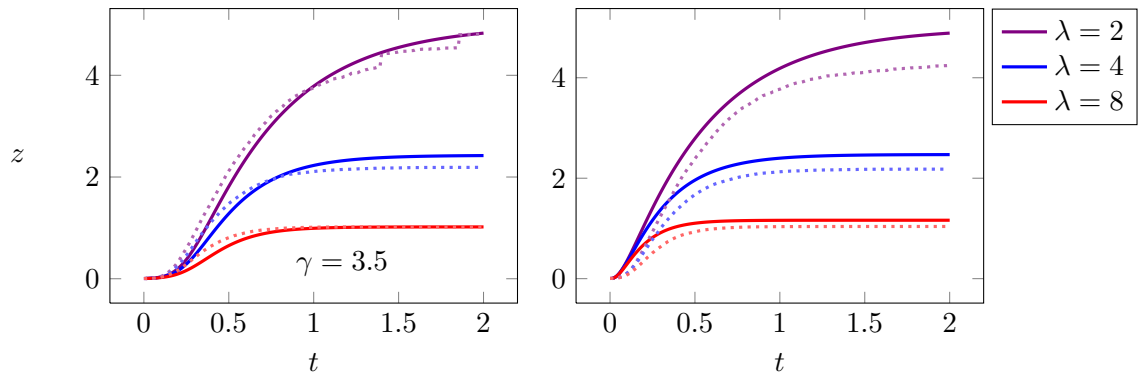

Figure \ref{fig1} compares the equation (\ref{dzdt}), the result of mean field theory, to numerical simulations.  We see that the qualitative sketch of the mean field trajectory is reproduced by the simulated dynamics for the range of $N$ tested, but quantitatively the curves appear shifted a bit, which is expected due to some of our approximations.  Interestingly, we see that for $\gamma=3.5$, the mean field theory slightly lags behind the simulations, whereas for $\gamma=2.5$, the mean field theory leads the simulated dynamics. This suggests, perhaps, that the sharp transition observed in mean field theory between $\gamma>3$ and $\gamma<3$ is likely not quite as sharp in the actual dynamics on a network.\footnote{Another issue is that $N=2000$ may be far too small to see a difference, but we did not have the computing power available to test this.}  

It is not hard to understand qualitatively what will happen if we assume that $\rho_k$ does not describe a scale free network.   In this case, we will no longer have an explicit form for the answer, but we can still understand the qualitative behavior by studying the quantity \begin{equation}
C(\gamma)\equiv \lim_{k\rightarrow\infty} \left[\frac{1}{k-m-1}\sum_{n=m}^k n^\gamma \rho_n \right].
\end{equation}We can use the divergences in $C(\gamma)$ to bound the dynamics on our given graph by replacing the graph's degree distribution with a non-normalized $\rho_k \sim k^{-\gamma}$, to find bounds in $\dot{z}\sim \theta(z)$.   Crudely speaking $C(\gamma) \sim k^\gamma \rho_k$ for large $k$, but to take care of a network where some of the $\rho_k$ may be 0, we will use the above definition.   If $C(3-\epsilon)=\infty$ for some $\epsilon>0$, then we conclude that $\tau_{\mathrm{spread}}\sim \mathrm{O}(1)$.   The case where $C(3)<\infty$ but $C(3+\epsilon)=\infty$ for any $\epsilon>0$ implies that $\tau_{\mathrm{spread}}\sim \log \log N$, which we obtain by bounding the spread time both from below and above by bounding $\theta(z)$ by two scale free distributions of degree $\gamma=3$.  If $C(3+\epsilon)<\infty$, then we conclude that $\tau_{\mathrm{spread}} \sim \log N$.  For this last case, there is an epidemic threshold independent of $N$, while for the former cases, there is only an epidemic threshold vanishing as $N\rightarrow\infty$.
\subsection{SIS Epidemic?}
A natural question to ask, given our success with mean field theory above, is whether or not we can do something for the SIS epidemic.    In the SIS epidemic, instead of dying (transitioning to state R), nodes transition to state S with rate $\lambda$.  The mean field equations in this case are given by \cite{pastor}:\begin{equation}
\dot{I}_k = k\theta (1-I_k) - \lambda I_k.
\end{equation}
Numerous problems arise in this case.   One of the major problems is that since it is possible to become susceptible again, we do not have the simple reduction of the S dynamics to a single equation.    The second, critical, problem is that $\dot{\theta}$ is not proportional to $\theta$ -- instead, we get a ``tower" of dynamical equations for the probability of looking at an infected node weighted by $k^2$, $k^3$, etc.   This implies that the irreversibility of the SIR epidemic is crucial for the exact solutions found above.
\subsection{STD Epidemics on Scale Free Bipartite Graphs}
A natural extension of the above discussion is the STD epidemic model on bipartite scale free graphs, as introduced in \cite{hetero}.\footnote{Actually, this paper considered the SIS epidemic.  But as we just mentioned, the SIS epidemic does not have a nice solution -- at least not using our techniques.}   The basic idea of this model is that there are two networks, a ``male" network and a ``female" network, such that all edges are between a male and female.   The mean field theory we used in the previous parts would be a bad approximation here, because we do have two distinct types of nodes, but at the expense of doubling the number of dynamical variables to $S_{\mathrm{M}k}$, $S_{\mathrm{F}k}$, $I_{\mathrm{M}k}$ and $I_{\mathrm{F}k}$, referring to the probability that a male/female node is susceptible and male/female node is infected respectively, we can correct for this.  For simplicity, let us assume that the male graph is scale free of degree $\gamma_{\mathrm{M}}$, and the female graph is scale free of degree $\gamma_{\mathrm{F}}$.    The extension of the mean field equations above is straightforward:\footnote{We assume that the rates are not M/F dependent, for simplicity, as was done in \cite{hetero}.} \begin{subequations}\begin{align}
\dot{S}_{\mathrm{F}k} &= -k\theta_{\mathrm{M}}S_{\mathrm{F}k}, \\
\dot{S}_{\mathrm{M}k} &= -k\theta_{\mathrm{F}}S_{\mathrm{M}k}, \\
\dot{I}_{\mathrm{F}k} &= k\theta_{\mathrm{M}}S_{\mathrm{F}k} - \lambda I_{\mathrm{F}k} ,\\
\dot{I}_{\mathrm{M}k} &= k\theta_{\mathrm{F}}S_{\mathrm{M}k} - \lambda I_{\mathrm{M}k}, 
\end{align}\end{subequations}with $\theta_{\mathrm{F}}$ and $\theta_{\mathrm{M}}$ defined in the same way as before:\begin{subequations}\begin{align}
\theta_{\mathrm{F}} &= \frac{1}{\langle k\rangle_{\mathrm{F}}} \sum k\rho_{\mathrm{F}k} I_{\mathrm{F}k} ,\\
\theta_{\mathrm{F}} &= \frac{1}{\langle k\rangle_{\mathrm{M}}} \sum k\rho_{\mathrm{M}k} I_{\mathrm{M}k}.
\end{align}\end{subequations}   By defining $z_{\mathrm{F}}$ and $z_{\mathrm{M}}$ as before: \begin{subequations}\begin{align}
z_{\mathrm{F}} &\equiv -\log S_{\mathrm{F}m}, \\
z_{\mathrm{M}} &\equiv -\log S_{\mathrm{M}m}, 
\end{align}\end{subequations}we find, using the same tricks as above,\begin{subequations}\begin{align}
\dot{z}_{\mathrm{F}} &= m \theta_{\mathrm{M}}, \\ 
\dot{z}_{\mathrm{M}} &= m\theta_{\mathrm{F}}, \\ 
\dot{\theta}_{\mathrm{F}} &= (\gamma_{\mathrm{F}}-2)mz^{\gamma_{\mathrm{F}}-3}\Gamma(3-\gamma_{\mathrm{F}},z_{\mathrm{F}})\theta_{\mathrm{M}} - \lambda\theta_{\mathrm{F}}, \\
\dot{\theta}_{\mathrm{M}} &= (\gamma_{\mathrm{M}}-2)mz^{\gamma_{\mathrm{M}}-3}\Gamma(3-\gamma_{\mathrm{M}},z_{\mathrm{M}})\theta_{\mathrm{F}} - \lambda\theta_{\mathrm{M}}.
\end{align}\end{subequations}

We have not found a way to solve these equations nearly exactly.  The difficulty comes in via the mixing of $\theta_{\mathrm{F}}$ and $\theta_{\mathrm{M}}$, which render the division trick we used earlier useless.   However, we can solve a simplified version of the model.   Consider the case where $\lambda=0$ -- this should be a decent approximation to the case where $\lambda \ll 1$ anyways (so the epidemic spreads very rapidly), and should give us qualitative insight into the nature of spreading.    In this case, we can once again employ the division trick, and we find that, just as before, using identities out of \cite{abramowitz}: \begin{subequations}\label{eq26ab}\begin{align}
\theta_{\mathrm{F}}(z_{\mathrm{F}}) &= z^{\gamma_{\mathrm{F}}-2}_{\mathrm{F}}\Gamma(3-\gamma_{\mathrm{F}},z_{\mathrm{F}}) + 1-\mathrm{e}^{-z_{\mathrm{F}}} = 1-(\gamma_{\mathrm{F}}-2)z^{\gamma_{\mathrm{F}}-2}\Gamma(2-\gamma_{\mathrm{F}},z), \\
\theta_{\mathrm{M}}(z_{\mathrm{M}}) &= z^{\gamma_{\mathrm{M}}-2}_{\mathrm{M}}\Gamma(3-\gamma_{\mathrm{M}},z_{\mathrm{M}}) + 1-\mathrm{e}^{-z_{\mathrm{M}}} = 1-(\gamma_{\mathrm{M}}-2)z^{\gamma_{\mathrm{M}}-2}\Gamma(2-\gamma_{\mathrm{M}},z).
\end{align}\end{subequations}Now, we use that \begin{equation}
\frac{\dot{z}_{\mathrm{F}}}{\dot{z}_{\mathrm{M}}} = \frac{\mathrm{d}z_{\mathrm{F}}}{\mathrm{d}z_{\mathrm{M}}} = \frac{\theta_{\mathrm{M}}}{\theta_{\mathrm{F}}} 
\end{equation}to find that \begin{equation}
F(z_{\mathrm{F}};\gamma_{\mathrm{F}}) = F(z_{\mathrm{M}};\gamma_{\mathrm{M}}) \label{fzf}
\end{equation} where \begin{equation}
F(z) = \int\limits_0^z \mathrm{d}z^\prime \theta(z^\prime).
\end{equation}
Returning to our assumption that the graphs are scale free:\begin{equation}
F(z;\gamma) \equiv z -\frac{\gamma-2}{\gamma-1}\left[1-\mathrm{e}^{-z} + z^{\gamma-1}\Gamma(2-\gamma,z)\right].
\end{equation}
Now, to understand (\ref{fzf}) in the regime of interest (for small $z$), we perform asymptotic expansions on $F$.  We find that the lowest order non vanishing terms are given by \begin{equation}
F(z;\gamma) \approx \left\lbrace\begin{array}{ll} \displaystyle \dfrac{\gamma-2}{2(\gamma-3)} z^2&\  \gamma > 3 \\ \displaystyle \dfrac{z^2}{2}\log\dfrac{1}{z} &\ \gamma=3 \\ \displaystyle \dfrac{\Gamma(3-\gamma)}{\gamma-1}z^{\gamma-1} &\ 2<\gamma<3 \end{array}\right..
\end{equation}

Let us look at a few examples of what this implies about the dynamics as the epidemic gets started.   Suppose that $\gamma_{\mathrm{F}}>3$ and $\gamma_{\mathrm{M}} > 3$.   It is easy to see that (\ref{fzf}) implies that \begin{equation}
z_{\mathrm{F}} \approx \sqrt{ \frac{(\gamma_{\mathrm{F}}-3)(\gamma_{\mathrm{M}}-2)}{(\gamma_{\mathrm{F}}-2)(\gamma_{\mathrm{M}}-3)}} z_{\mathrm{M}},  \label{zfzm1}
\end{equation}or\begin{equation}
S_{\mathrm{F}k} \approx S_{\mathrm{M}k}^{\sqrt{(\gamma_{\mathrm{F}}-3)(\gamma_{\mathrm{M}}-2)/(\gamma_{\mathrm{F}}-2)(\gamma_{\mathrm{M}}-3)}}.
\end{equation}We should not take the precise exponent here particularly seriously, but just note that the fraction of male susceptible nodes is some power of the fraction of female susceptible nodes.   Now, let us consider the case where $\gamma_{\mathrm{F}}>3$ but $\gamma_{\mathrm{M}}<3$.   Then we find \begin{equation}
z_{\mathrm{F}} = \sqrt{\frac{2(\gamma_{\mathrm{F}}-3)\Gamma(3-\gamma_{\mathrm{M}})}{(\gamma_{\mathrm{F}}-2)(\gamma_{\mathrm{M}}-1)}}z_{\mathrm{M}}^{(\gamma_{\mathrm{M}}-1)/2}.  \label{zfzm2}
\end{equation}This is a surprising result -- for very small $t$, the female nodes get infected at a rate more than exponentially faster than to the male nodes, although this range of times is not very long.  

We can also see quickly that a similar result for $\tau_{\mathrm{spread}}$ holds:   if $\gamma_{\mathrm{M}},\gamma_{\mathrm{F}} >3$, the spreading  dynamics are $\mathrm{O}(\log N)$; they are $\mathrm{O}(1)$ in the case of $\gamma_{\mathrm{M}}<3$.  In the case of $\gamma_{\mathrm{M}},\gamma_{\mathrm{F}}>3$, this follows from (\ref{zfzm1}) and (\ref{eq26ab}): \begin{equation}
\dot{z}_{\mathrm{F}} = m\theta_{\mathrm{M}} \approx m\frac{\gamma_{\mathrm{M}}-2}{\gamma_{\mathrm{M}}-3}z_{\mathrm{M}} = m\sqrt{\frac{(\gamma_{\mathrm{M}}-2)(\gamma_{\mathrm{F}}-2)}{(\gamma_{\mathrm{M}}-3)(\gamma_{\mathrm{F}}-3)}}z_{\mathrm{F}}.
\end{equation}and similarly for $z_{\mathrm{M}}$.   In the case of $\gamma_{\mathrm{M}}<3$, $\gamma_{\mathrm{F}}>3$, we have instead, using (\ref{zfzm2}) and (\ref{eq26ab}):\begin{equation}
\dot{z}_{\mathrm{F}} \sim \theta_{\mathrm{M}} \sim z_{\mathrm{M}}^{\gamma_{\mathrm{M}}-2} \sim z_{\mathrm{F}}^{2(\gamma_{\mathrm{M}}-2)/(\gamma_{\mathrm{M}}-1)}.
\end{equation}Since \begin{equation}
0 < 2\frac{\gamma_{\mathrm{M}}-2}{\gamma_{\mathrm{M}}-1} < 1 \;\;\;\;\; (2<\gamma_{\mathrm{M}}<3)
\end{equation}we conclude that growth is faster than linear, and that the spreading dynamics is O(1) for the same reason as in the SIR epidemic.    In the case of $\gamma_{\mathrm{F}}>3$, $\gamma_{\mathrm{M}}=3$, we find that since $z_{\mathrm{F}}^2 \sim -z_{\mathrm{M}}^2 \log z_{\mathrm{M}}$, that \begin{equation}
\dot{z}_{\mathrm{F}} = z_{\mathrm{M}} \log \frac{1}{z_{\mathrm{M}}} \approx z_{\mathrm{F}} \sqrt{\log \frac{1}{z_{\mathrm{F}}} + \mathrm{O}(\log\log z_{\mathrm{F}})}
\end{equation} Defining $y_{\mathrm{F}}=-\log z_{\mathrm{F}}$ as we did earlier, we find that \begin{equation}
\tau_{\mathrm{spread}} \sim\int \frac{\mathrm{d}y_{\mathrm{F}}}{\sqrt{y_{\mathrm{F}}}} \sim \sqrt{\log N}
\end{equation} In the case of $\gamma_{\mathrm{M}}=\gamma_{\mathrm{F}}=3$, we can find that $\tau_{\mathrm{spread}}\sim \log\log N$ just as before.

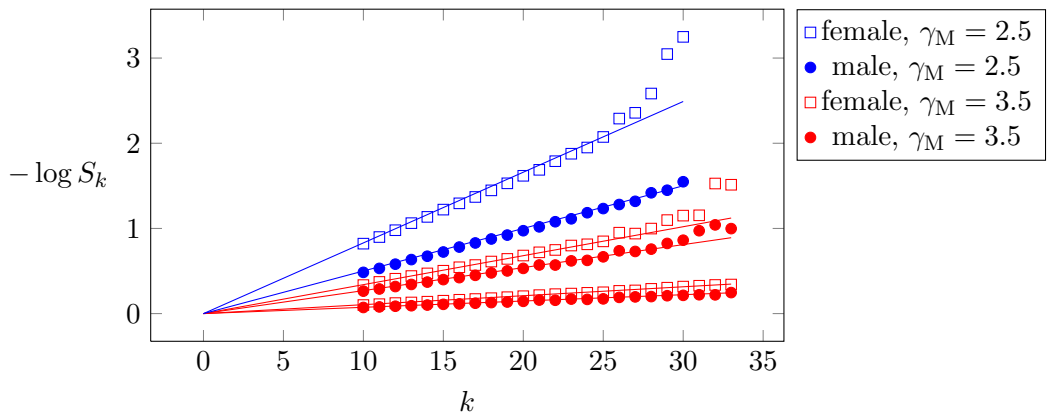
\begin{figure}[here]
\centering
\begin{tikzpicture}

\begin{axis}[width=10cm, height=6cm, xlabel=$k$, ylabel=$-\log S_k$, ylabel style=sloped like x axis]
\pgfplotstableread{STD25.txt}\datatable
\addplot[color=blue, mark=square, only marks] table[x index=0, y index=12] from \datatable;
\addlegendentry{female, $\gamma_{\mathrm{M}}=2.5$}
\addplot[color=blue, mark=*, only marks] table[x index=0, y index=11] from \datatable;
\addlegendentry{male, $\gamma_{\mathrm{M}}=2.5$}

\pgfplotstableread{STD35.txt}\datatable
\addplot[color=red, mark=square, only marks] table[x index=0, y index=12] from \datatable;
\addlegendentry{female, $\gamma_{\mathrm{M}}=3.5$}

\addplot[color=red, mark=*, only marks] table[x index=0, y index=11] from \datatable;
\addlegendentry{male, $\gamma_{\mathrm{M}}=3.5$}
\addplot[color=red, domain=0:33] {0.027*x};
\addplot[color=red, domain=0:33] {0.0105*x};

\addplot[color=red, mark=square, only marks] table[x index=0, y index=16] from \datatable;
\addplot[color=red, mark=*, only marks] table[x index=0, y index=15] from \datatable;
\addplot[color=red, domain=0:33] {0.0074*x};
\addplot[color=red, domain=0:33] {0.034*x};

\addplot[color=blue, domain=0:30] {0.083*x};
\addplot[color=blue, domain=0:30] {0.05*x};
\end{axis}
\end{tikzpicture}
\caption{$-\log S_{\mathrm{M}k}$  and $-\log S_{\mathrm{F}k}$ on SI STD epidemics on graphs with $N=5000$ nodes and $\gamma_{\mathrm{F}}=3.5$, averaged over 400 trials.   We used times $t=0.24$ and $0.32$ for $\gamma_{\mathrm{M}}=3.5$, and $0.24$ for $\gamma_{\mathrm{M}}=2.5$, to avoid finite size effects (which become visible for the blue lines), as discussed earlier.   We have checked that other parameters lead to similar linear relations.}
\label{zexp}
\end{figure}

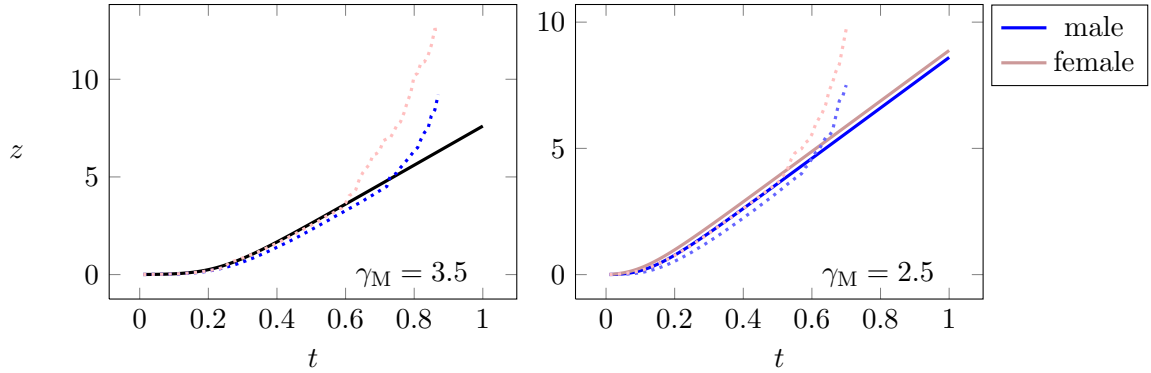
\begin{figure}[here]
\centering
\begin{tikzpicture}
\begin{axis}[width=7cm, height=5.5cm, xlabel=$t$, ylabel=$z$, ylabel style=sloped like x axis]

\pgfplotstableread{nu35rec.txt}\datatable
\addplot[color=black, very thick] table[x index=0, y index=1] from \datatable;

\pgfplotstableread{testSTD35.txt}\datatable
\addplot[color=blue, very thick, dotted] table[x index=0, y index=1] from \datatable;
\addplot[color=pink, very thick, dotted] table[x index=0, y index=2] from \datatable;

\draw (axis cs: 0.6, 0) node[right] {$\gamma_{\mathrm{M}}=3.5$};
\end{axis}
\begin{scope}[xshift=6.2cm]
\begin{axis}[width=7cm, height=5.5cm, xlabel=$t$]

\pgfplotstableread{nu25stdM.txt}\datatable
\addplot[color=blue, very thick] table[x index=0, y index=1] from \datatable;
\addlegendentry{male}
\pgfplotstableread{nu25stdF.txt}\datatable
\addplot[color=pink!80!black, very thick] table[x index=0, y index=1] from \datatable;
\addlegendentry{female}

\pgfplotstableread{testSTD25.txt}\datatable
\addplot[color=blue!60!white, very thick, dotted] table[x index=0, y index=1] from \datatable;
\addplot[color=pink, very thick, dotted] table[x index=0, y index=2] from \datatable;

\draw (axis cs: 0.6, 0) node[right] {$\gamma_{\mathrm{M}}=2.5$};

\end{axis}
\end{scope}
\end{tikzpicture}
\caption{Comparison of $z(t)$ between theory (solid line) and simulations (dotted line) for the SI STD epidemic model.  We used $N=2000$, $m=10$, and 100 trials.}
\label{zexp5}
\end{figure}

To generate bipartite scale free networks for use in simulations, we used a similar algorithm to what is used in \cite{hetero}, which unfortunately does not guarantee that all F nodes have at least 10 edges.   However, we see that this does not significantly ruin the dynamics, and they match mean field theory extremely well, as shown in Figure \ref{zexp5}, although they are a bit lower than mean field theory would predict in the range of validity.   Figure \ref{zexp} shows that the fraction of susceptible nodes (for both M and F) is exponentially decaying with $k$, as mean field theory predicts.   Together, these suggest that mean field theory is a valid dynamical approximation at all times, notwithstanding finite size limitations.

\section{Rumor Spreading} \label{rusec}
Now, let us turn the discussion to models of rumor spreading.   The essential idea of the rumor spreading model is that people can be described as either unaware of the rumor (state S), actively spreading the rumor (state I), and not actively spreading the rumor, and having heard of it (state R).   The key difference with the SIR epidemic is that the death rates will now change.  

There are 2 possibilities.  The classic rumor spreading model, which we will denote ``type IR" rumor spreading, corresponds to a situation where every edge that connects a given node in state I to a state in either I or R induces transitions to R with rate $\lambda$.   We will instead consider a simplified version, which we denote ``type I" rumor spreading, where only I nodes induce such transitions.   Type I rumor spreading is perhaps not as realistic as type IR, for dynamical reasons which will become clear, but it will admit an exact solution of the same type as we have found before, so we will focus our discussion on this model.  First, we begin by discussing type IR rumor spreading, and describe what can be obtained from mean field theory.
\subsection{Type IR Rumor Spreading}
Let us define\begin{equation}
\psi = \sum \frac{k\rho_k S_k}{\langle k\rangle}.
\end{equation} The mean field equations are\begin{subequations}\begin{align}
\dot{S}_k &= -k\theta S_k, \\ 
\dot{I}_k &= k\theta S_k - \lambda k (1-\psi) I_k.  \label{ik2}
\end{align}\end{subequations}
We will not find a way to nearly exactly solve the above equations, even for a scale free graph.   Furthermore, essentially all of the results we find in this section can be found in \cite{nekovee}, but we repeat them here for completeness, and because we derive them in a slightly quicker way.    We begin by noting that introducing $z$ as we did before, we find the exact same relation that $S_k = \mathrm{e}^{-kz/m}$.  In particular, this means that (once again, for simplicity, assuming $\rho_k\sim k^{-\gamma}$)\begin{equation}
\psi(z) = \sum \frac{k\rho_k}{\langle k\rangle} S_k = \sum \frac{k\rho_k}{\langle k\rangle} \mathrm{e}^{-kz/m} \approx \frac{\gamma -1}{\langle k\rangle} \int\limits_m^\infty \mathrm{d}k \left(\frac{m}{k}\right)^{\gamma-1} \mathrm{e}^{-kz/m} =  (\gamma-2)z^{2-\gamma}\Gamma(2-\gamma,z).
\end{equation}In general, we can find an expression for $\psi(z)$ for more complicated degree distributions, but we may not be able to find the exact solution.  Given $\psi(z)$, (\ref{ik2}) becomes \begin{equation}
\dot{I}_k = k\theta \mathrm{e}^{-kz/m} - \lambda k(1-\psi(z))I_k.
\end{equation}

Unfortunately, it is far from obvious how to solve these differential equations exactly.   Although they are linear in $I$, they involve diagonalizing a nontrivial infinite dimensional matrix.   We will content ourselves to merely understanding the location of the fixed point $z^*$.   To find $z^*$, we note that \begin{equation}
\frac{\mathrm{d}}{\mathrm{d}t} \langle I\rangle = \sum \rho_k\dot{I}_k = \sum k\rho_k \mathrm{e}^{-kz/m}\theta - \lambda(1-\psi(z))\sum k\rho_k I_k = \langle k\rangle [ \theta \psi(z) - \lambda\theta(1-\psi(z))].
\end{equation}At $t\rightarrow\infty$, this should go to 0, so we conclude that \begin{equation}
\psi(z^*) = \frac{\lambda}{\lambda+1}.
\end{equation}
We can say more about the state of the graph at the fixed point: the mean field theory clearly predicts that $S_k(\infty)$ decreases exponentially with $k$.   This fact was known to \cite{moreno2}, but a theoretical reason was not known.

Since the focus of this paper is on discovering exact solutions, let us now turn to type I rumor spreading, which we will discover does have an exact solution.

\subsection{Type I Rumor Spreading}
Let us now turn to the simplified model of type I rumor spreading, with mean field equations \begin{subequations}\begin{align}
\dot{S}_k &= -k\theta S_k, \\
\dot{I}_k &= k\theta S_k - \lambda k\theta I_k.
\end{align}\end{subequations}It is clear that $S_k = \mathrm{e}^{-kz/m}$ as before.   We now may exploit a different trick than the one we have previously used.   Consider \begin{equation}
\frac{\dot{I}_k}{\dot{S}_k} = \frac{\mathrm{d}I_k}{\mathrm{d}S_k} = -1 + \lambda \frac{I_k}{S_k}.  \label{eq26}
\end{equation}Then we see that by defining \begin{equation}
w_k S_k = I_k,
\end{equation}(\ref{eq26}) becomes, assuming for simplicity that $\lambda\ne 1$,\footnote{The case of $\lambda=1$ is not difficult to solve, but we do not present it in this paper.} \begin{equation}
 S_k \frac{\mathrm{d}w_k}{\mathrm{d}S_k} = -1 + (\lambda-1)w_k,
\end{equation}which for appropriate initial conditions, implies \begin{equation}
\frac{1}{\lambda-1} \log \frac{(\lambda-1)w_k - 1}{-1} = \log S_k,
\end{equation}or \begin{equation}
I_k = \frac{1}{1-\lambda} \left(S_k^\lambda - S_k\right) = \frac{\mathrm{e}^{-\lambda kz/m} - \mathrm{e}^{-kz/m}}{1-\lambda}.  \label{iksk}
\end{equation}
Now, from here, we can directly compute $\theta(z)$.  As we expect, $\theta(z)$ has an explicit expression for a scale free graph under the sum to integral approximation: \begin{equation}
\theta(z) \approx \int\limits_m^\infty \frac{\gamma-2}{m} \mathrm{d}k \left(\frac{m}{k}\right)^{\gamma-1}  \frac{\mathrm{e}^{-\lambda kz/m} - \mathrm{e}^{-kz/m}}{1-\lambda} = \frac{\gamma-2}{1-\lambda}\left[(\lambda z)^{\gamma-2}\Gamma(2-\gamma,\lambda z) - z^{\gamma-2}\Gamma(2-\gamma,z)\right]
\end{equation}and therefore obtain \begin{equation}
\dot{z} = m\theta =  \frac{\gamma-2}{1-\lambda} m \left[(\lambda z)^{\gamma-2}\Gamma(2-\gamma,\lambda z) - z^{\gamma-2}\Gamma(2-\gamma,z)\right].
\end{equation}Using $\Gamma$ function identities we can re-write this expression: \begin{equation}
\dot{z} = m\frac{\mathrm{e}^{-\lambda z} - \mathrm{e}^{-z} - (\lambda z)^{\gamma-2} \Gamma(3-\gamma,\lambda z)+z^{\gamma-2}\Gamma(3-\gamma,z)}{1-\lambda}
\end{equation}
Just as before, we can find the exact solution by finding $t$ in terms of $z$, expressed as an integral.   Interestingly, we should note that for type I rumor spreading it is actually far easier to extract the relevant physical information: $S_k$ and $I_k$, than for the SIR epidemic.   Determining $S_k$ is the same as for the epidemics, but this time we can simply read off $I_k$ from (\ref{iksk}).

Let's analyze the behavior of this equation for small $z$.   When $\gamma>3$, we use the asymptotic expansions for $z\approx 0$: \begin{equation}
\dot{z}\approx m\frac{-\lambda z + z -(3-\gamma)^{-1}(\lambda z - z)}{1-\lambda} = \frac{\gamma-2}{\gamma -3 }mz,
\end{equation}which is precisely what we would have found had we naively assumed that the short time behavior of the rumor spreading was behaving like a SIR epidemic with effective death rate of 0.  Our intuition thus implies that we should have expected the absence of an epidemic threshold, which is indeed what we see.    However, the intuition of approximating rumor spreading as an epidemic fails for the case of $\gamma <3$, interestingly, where the dominant asymptotic behavior near the origin comes exclusively from the $\Gamma$ functions: \begin{equation}
\dot{z}\approx m\Gamma(3-\gamma)\frac{1-\lambda^{\gamma-2}}{1-\lambda} z^{\gamma-2}. \label{lambdab1}
\end{equation}Here, interestingly, we see that the death rate has an effect on the short time dynamics even for small $z$:  the $\lambda$ dependent factor behaves like $1$ for $\lambda \ll 1$, and $\lambda^{-(3-\gamma)}$ for $\lambda\gg 1$ (as expected, higher death rates suppress the growth of the epidemic).   We also note that it is obvious from here that $\tau_{\mathrm{spread}}$ has the same scaling behavior as with the SIR epidemic:  $\mathrm{O}(\log N)$ when $\gamma>3$, $\mathrm{O}(\log\log N)$ when $\gamma=3$, and $\mathrm{O}(1)$ for $\gamma< 3$.

\begin{figure}[here]
\centering
\begin{tikzpicture}
\begin{axis}[width=7cm, height=5.5cm, xlabel=$t$, ylabel=$z$, ylabel style=sloped like x axis]

\pgfplotstableread{2nu354.txt}\datatable
\addplot[color=red, very thick] table[x index=0, y index=1] from \datatable;

\pgfplotstableread{2nu352.txt}\datatable
\addplot[color=orange, very thick] table[x index=0, y index=1] from \datatable;

\pgfplotstableread{2nu3505.txt}\datatable
\addplot[color=blue, very thick] table[x index=0, y index=1] from \datatable;

\pgfplotstableread{2nu35025.txt}\datatable
\addplot[color=violet, very thick] table[x index=0, y index=1] from \datatable;

\pgfplotstableread{test354.txt}\datatable
\addplot[color=red!60!white, very thick, dotted] table[x index=0, y index=1] from \datatable;

\pgfplotstableread{test352.txt}\datatable
\addplot[color=orange!60!white, very thick, dotted] table[x index=0, y index=1] from \datatable;

\pgfplotstableread{test3505.txt}\datatable
\addplot[color=blue!60!white, very thick, dotted] table[x index=0, y index=1] from \datatable;

\pgfplotstableread{test35025.txt}\datatable
\addplot[color=violet!60!white, very thick, dotted] table[x index=0, y index=1] from \datatable;

\draw (axis cs: 0, 4) node[right] {$\gamma=3.5$};
\end{axis}
\begin{scope}[xshift=6.2cm]
\begin{axis}[width=7cm, height=5.5cm, xlabel=$t$]
\pgfplotstableread{2nu25l4.txt}\datatable
\addplot[color=red, very thick] table[x index=0, y index=1] from \datatable;
\addlegendentry{$\lambda=4$};

\pgfplotstableread{2nu25l2.txt}\datatable
\addplot[color=orange, very thick] table[x index=0, y index=1] from \datatable;
\addlegendentry{$\lambda=2$};

\pgfplotstableread{2nu25l05.txt}\datatable
\addplot[color=blue, very thick] table[x index=0, y index=1] from \datatable;
\addlegendentry{$\lambda=0.5$};

\pgfplotstableread{2nu25l025.txt}\datatable
\addplot[color=violet, very thick] table[x index=0, y index=1] from \datatable;
\addlegendentry{$\lambda=0.25$};

\pgfplotstableread{test254.txt}\datatable
\addplot[color=red!60!white, very thick, dotted] table[x index=0, y index=1] from \datatable;

\pgfplotstableread{test252.txt}\datatable
\addplot[color=orange!60!white, very thick, dotted] table[x index=0, y index=1] from \datatable;

\pgfplotstableread{test2505.txt}\datatable
\addplot[color=blue!60!white, very thick, dotted] table[x index=0, y index=1] from \datatable;

\pgfplotstableread{test25025.txt}\datatable
\addplot[color=violet!60!white, very thick, dotted] table[x index=0, y index=1] from \datatable;

\draw (axis cs: 0, 4) node[right] {$\gamma=2.5$};

\end{axis}
\end{scope}
\end{tikzpicture}
\caption{Comparison of $z(t)$ between theory (solid line) and simulations (dotted line) for type I rumor spreading.  We used $N=2000$, $m=10$ and averaged over 50 trials.  It required a time step of $\Delta t \approx 0.01$ before the simulation appeared to accurately reflect continuous time dynamics.}
\label{zexp3}
\end{figure}
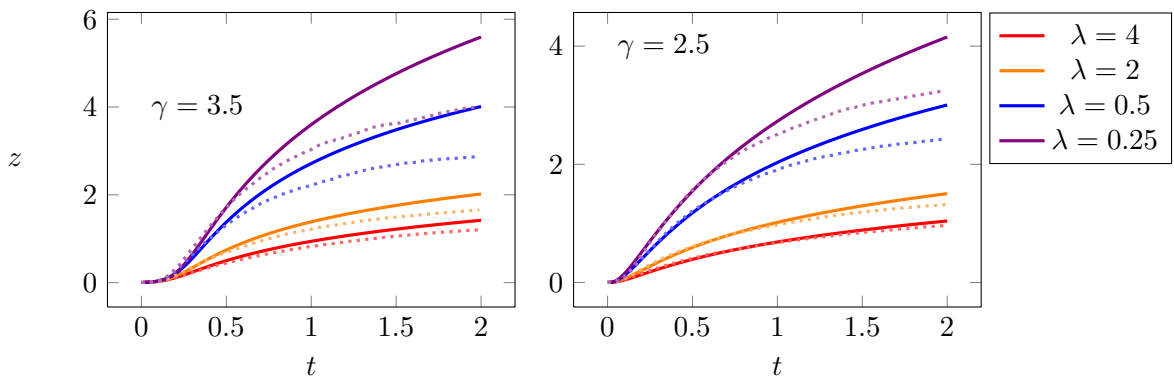

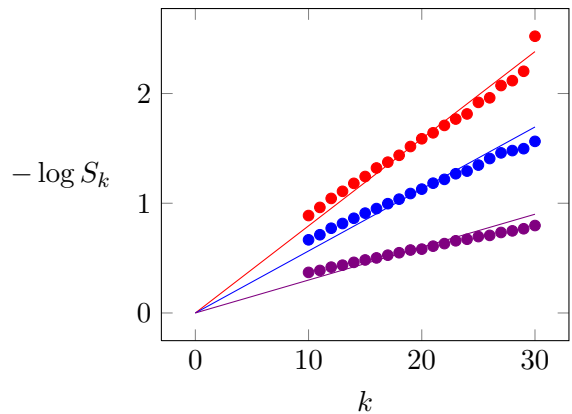
\begin{figure}[here]
\centering
\begin{tikzpicture}

\begin{axis}[width=7cm, height=6cm, xlabel=$k$, ylabel=$-\log S_k \;\;$, ylabel style=sloped like x axis]

\pgfplotstableread{Rnu35n5000z2.txt}\datatable
\addplot[color=violet, mark=*, only marks] table[x index=0, y index=10] from \datatable;
\addplot[color=blue, mark=*, only marks] table[x index=0, y index=17] from \datatable;
\addplot[color=red, mark=*, only marks] table[x index=0, y index=25] from \datatable;

\addplot[color=red, domain=0:30] {0.0794*x};
\addplot[color=blue, domain=0:30] {0.0565*x};
\addplot[color=violet, domain=0:30] {0.03*x};
\end{axis}

\end{tikzpicture}
\caption{$-\log s_k$ as a function of $k$ at various times.   Here we show the example of growth on a scale free graph of degree $\gamma=3.5$ with $N=5000$ nodes and death rate $\lambda=4$, averaged over 200 trials.}
\label{zexp2}
\end{figure}

Figure \ref{zexp3} shows plots of the simulated rumor spreading, compared to mean field theory.   We see that at initial times, mean field theory is an excellent approximation, although it begins to significantly break down at large $z$.   The reason for this will be explained in the next subsection.    Figure \ref{zexp2} shows that $S_k$ is still exponentially decaying with $k$ for the rumor spreading models.   While for earlier times, the optimal linear fit requires a nonzero intercept with the $z$ axis, the qualitative picture of mean field theory holds very well.   

\subsection{Late Time Type I Dynamics}
The above discussion focuses on the early time dynamics.   For late times, we will see that type I rumor spreading is a simple example of a process where we should expect mean field theory to completely break down, something which we observed in Figure \ref{zexp3}.

Let us begin by naively assuming that mean field theory is an accurate description, and see what we find.  Proceeding as before:\begin{equation}
\dot{z} \approx \frac{\gamma-2}{1-\lambda}m\left[\frac{\mathrm{e}^{-\lambda z}}{\lambda z} - \frac{\mathrm{e}^{-z}}{z}\right] \approx \frac{\gamma-2}{|1-\lambda|m} \frac{\mathrm{e}^{-\min(1,\lambda)z}}{\min(1,\lambda)z}
\end{equation}
This implies that, letting $\Lambda=\min(1,\lambda)$, \begin{equation}
\tau_{\mathrm{end}} \sim \int\limits_{\mathrm{O}(1)}^{z^*}\mathrm{d}z \; \Lambda z \mathrm{e}^{\Lambda z} \sim \Lambda z^* \mathrm{e}^{\Lambda z^*}.
\end{equation}Now, we have to be careful about $z^*$.   In the type I rumor spreading, once all of an infected node's neighbors die, he will stay infected forever.   Suppose we are on a fully connected graph -- then it is clear that $z^* = -\log(1/N) = \log N$, and thus \begin{equation}
\tau_{\mathrm{end}} \sim N^\Lambda \log N.
\end{equation}
This is a very interesting and strange result -- the time scale itself of the epidemic ending is extremely sensitive on the parameters of the problem, until the critical point when $\lambda= 1$, in which case, roughly speaking, the epidemic spreads by pairs becoming infected, with one of the two quickly dying off.

This expression for $\tau_{\mathrm{end}}$ is completely incorrect, however, for a graph which is not fully connected.   Here, it becomes a little bit subtle to determine the correct $z^*$.   The basic intuition we have proceeds as follows.   Typically, the more connected a node was, the more likely it was to have gotten infected early, and to have died quickly.   Therefore, the nodes which survive are the ones with fewer connections.  Now, let us consider for simplicity, only the nodes which have on the order of the fewest connections, $m$.   If we choose a node to ``live" and kill all of its neighbors, repeating this process until we have saved or killed all nodes, then, since we expect to kill $\sim m$ nodes each time, we should expect that $s_m \sim m^{-1}$, or $z^*\sim \log m$.   

\begin{figure}[here]
\centering
\begin{tikzpicture}
\begin{semilogxaxis}[width=10cm, height=7cm, xlabel=$N$ (logarithmic plot), xtick={100, 200, 400, 800, 1600, 3200, 6400}, xticklabels={100, 200, 400, 800, 1600, 3200, 6400}, ylabel=$\tau_{\mathrm{end}}$, ylabel style=sloped like x axis]
\addplot[color=red, only marks, mark=*] coordinates {(100, 3.4104) (200, 4.05) (400, 5.01) (800, 5.61) (1600, 6.11) (3200, 6.85) (6400, 7.83)};
\addlegendentry{$\lambda=0.3$};
\addplot[color=orange, only marks, mark=*] coordinates {(100, 2.13) (200, 2.97)  (400, 3.07) (800, 3.6) (1600, 3.843) (3200, 4.41) (6400, 4.8)};
\addlegendentry{$\lambda=0.6$};
\addplot[color=green, only marks, mark=*] coordinates {  (100, 1.761) (200, 2.232) (400, 2.58) (800,  3.33) (1600, 3.55) (3200, 4.26) (6400, 4.68)};
\addlegendentry{$\lambda=1$};
\addplot[color=blue, only marks, mark=*] coordinates {(100, 2.147) (200, 2.52) (400, 3.15) (800, 3.87) (1600, 3.96) (3200, 4.83) (6400, 5.25) };
\addlegendentry{$\lambda=2$};
\addplot[color=violet, only marks, mark=*] coordinates { (100, 2.76) (200, 3.66)  (400, 3.93) (800, 4.5) (1600, 5.46) (3200, 5.97) (6400,  6.9)};
\addlegendentry{$\lambda=5$};

\addplot[color=red, domain=100:6400] {1.03*ln(x)-1.32};
\addplot[color=orange, domain=100:6400] {0.6*ln(x)-0.47};

\addplot[color=green, domain=100:6400] {0.71*ln(x)-1.54};
\addplot[color=blue, domain=100:6400] {0.76*ln(x)-1.4};

\addplot[color=violet, domain=100:6400] {0.96*ln(x)-1.655};
\end{semilogxaxis}
\end{tikzpicture}
\caption{The ending time, averaged over 50 trials, on scale free graphs with $\gamma=3.5$.  We can see that $t_{\mathrm{end}} \sim \log N$.   To speed up simulations, we used fairly large time steps -- we do not think this should alter the qualitative nature of the end time dynamics, although this may make our simulated $\tau_{\mathrm{end}}$ too small.}
\label{endt}
\end{figure}
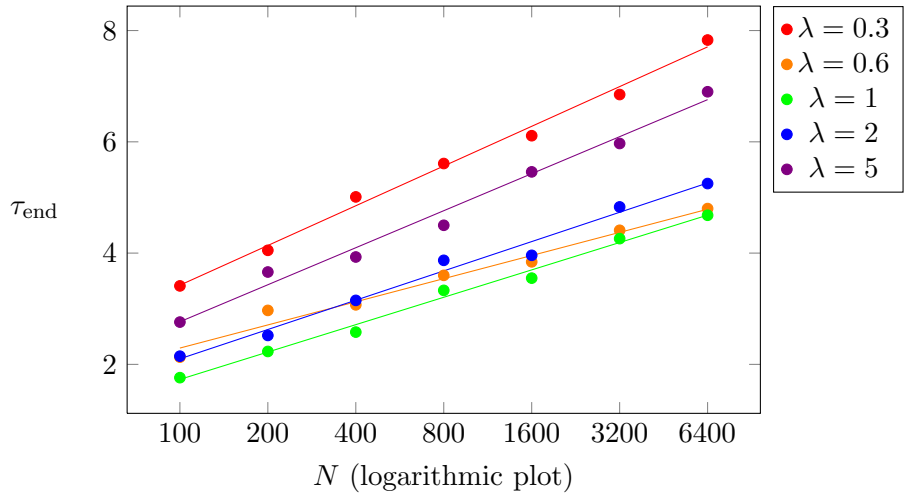

However, if the dynamics is driven to a fixed point at $z^*\sim \log m$, then we know that the mean field theory description must have completely broken down, since there is no fixed point for finite $z$.  The naive guess is that since the fixed point occurs at $z^*=\mathrm{O}(1)$, the fixed point is absolutely stable, and therefore $\tau_{\mathrm{end}}\sim \log N$.  We can qualitatively see this result holds up against numerical simulations, shown in Figure \ref{endt}.   Interestingly, we see that the dynamics ends fastest when $\lambda\approx 1$, and becomes slower both for large and small $\lambda$.  This has an intuitive interpretation -- for $\lambda\ll 1$, the ending dynamics is slow because we are waiting for death events, which take a very long time;  for $\lambda\gg 1$, the ending dynamics is slow because deaths occur so fast that the rumor/infection must propagate ``one node at a time" with a creation of an I-I edge quickly followed by one of the two dying.

\section{Recommendation Spreading}\label{sad}
We now show that a very recently proposed model for recommendation in social systems \cite{blattner} also has an exact solution in terms of an integral, just as we found above.   In this model, there are 3 states:  a susceptible node (S), an accepting node (A), and a denying node (D).   Instead of SIR-type dynamics, the dynamics of this model are as follows:  if an S comes in contact with an A, it will transition to an A with rate 1, and a D with rate $\lambda$.   This occurs per edge, so the mean field equations are \begin{equation}
D_k = 1-A_k-S_k
\end{equation}using conservation of probability, and \begin{subequations}\begin{align}
\dot{S}_k &= -(1+\lambda)k\theta S_k, \\
\dot{A}_k &= k\theta S_k.
\end{align}\end{subequations}Here we are using $A_k$ and $D_k$ for the fraction of nodes with $k$ edges in states A and D, respectively.   From our above work, it is clear that these equations have an exact solution in terms of an integral.

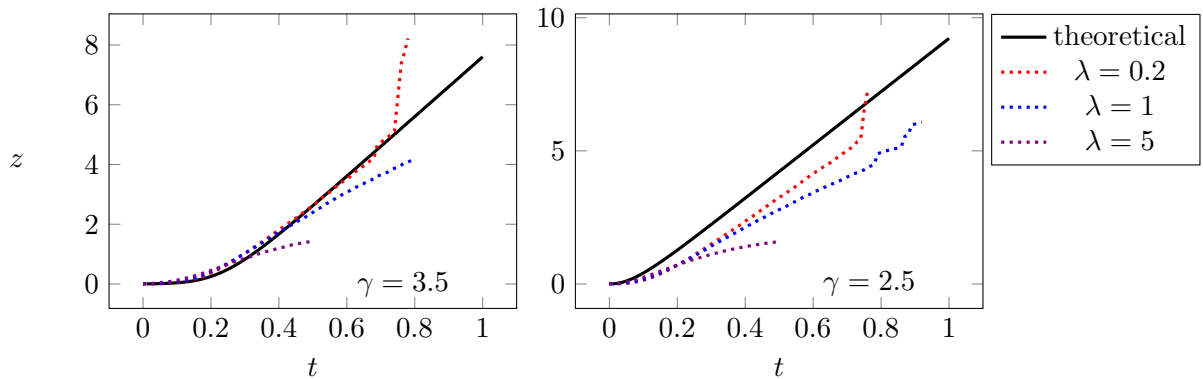
\begin{figure}[here]
\centering
\begin{tikzpicture}
\begin{axis}[width=7cm, height=5.5cm, xlabel=$t$, ylabel=$z$, ylabel style=sloped like x axis]

\pgfplotstableread{nu35rec.txt}\datatable
\addplot[color=black, very thick] table[x index=0, y index=1] from \datatable;

\pgfplotstableread{testrec02.txt}\datatable
\addplot[color=red, very thick, dotted] table[x index=0, y index=1] from \datatable;
\pgfplotstableread{testrec1.txt}\datatable
\addplot[color=blue, very thick, dotted] table[x index=0, y index=1] from \datatable;
\pgfplotstableread{testrec5.txt}\datatable
\addplot[color=violet, very thick, dotted] table[x index=0, y index=1] from \datatable;

\draw (axis cs: 0.6, 0) node[right] {$\gamma=3.5$};
\end{axis}
\begin{scope}[xshift=6.2cm]
\begin{axis}[width=7cm, height=5.5cm, xlabel=$t$]

\pgfplotstableread{nu25rec.txt}\datatable
\addplot[color=black, very thick] table[x index=0, y index=1] from \datatable;
\addlegendentry{theoretical};

\pgfplotstableread{testrec022.txt}\datatable
\addplot[color=red, very thick, dotted] table[x index=0, y index=1] from \datatable;
\addlegendentry{$\lambda=0.2$};
\pgfplotstableread{testrec12.txt}\datatable
\addplot[color=blue, very thick, dotted] table[x index=0, y index=1] from \datatable;
\addlegendentry{$\lambda=1$};
\pgfplotstableread{testrec52.txt}\datatable
\addplot[color=violet, very thick, dotted] table[x index=0, y index=1] from \datatable;
\addlegendentry{$\lambda=5$};

\draw (axis cs: 0.6, 0) node[right] {$\gamma=2.5$};

\end{axis}
\end{scope}
\end{tikzpicture}
\caption{Comparison of $z(t)$ between theory (solid line) and simulations (dotted line) for the recommendation spreading model.  We used $N=2000$, $m=10$, and 50 trials.   The significant deviations from mean field theory for $z$ suddenly increasing are due to finite size.  The deviations for $z$ flattening out are due to the breakdown of mean field theory discussed below.   We have cut off the trajectories once they begin to show significant deviations.}
\label{zexp4}
\end{figure}

For simplicity, let us focus on the case of a scale free graph.   We find that \begin{equation}
\frac{\dot{S}_k}{\dot{A}_k} = \frac{\mathrm{d}S_k}{\mathrm{d}A_k} = -(1+\lambda),
\end{equation}which implies that \begin{equation}
A_k = \frac{1-S_k}{1+\lambda}.
\end{equation}This implies that, to good approximation, using $z$ as defined above: \begin{equation}
\theta = \frac{1}{1+\lambda} \frac{1}{\langle k\rangle} \sum k\rho_k A_k = \frac{1-(\gamma-2)z^{\gamma-2}\Gamma(2-\gamma,z)}{1+\lambda} = \frac{z^{\gamma-2}\Gamma(3-\gamma,z)+1-\mathrm{e}^{-z}}{1+\lambda}.
\end{equation}
We immediately see that \begin{equation}
\dot{z} = m\left[z^{\gamma-2}\Gamma(3-\gamma,z)+1-\mathrm{e}^{-z}\right].   \label{dotz2}
\end{equation}
At mean field level, we recognize this as exactly the same as SI epidemic dynamics.   This is not an accident, and we will explain why this occurs shortly.  Our previous analysis implies that $\tau_{\mathrm{spread}}\sim \log N$ if $\gamma>3$, $\sim \log \log N$ if $\gamma=3$ and $\sim \mathrm{O}(1)$ for $\gamma < 3$.     In this case, for large $z$, the dominant term in the dynamics is actually the term 1, so we conclude that $\tau_{\mathrm{end}}\sim \log N$ for this model.   Figure \ref{zexp4} compares the theoretical dynamics of this model to mean field theory, where we see excellent agreement for $\gamma=3.5$ (for short times, at least) and qualitative agreement for $\gamma=2.5$, but with the simulated $z$ a bit smaller than theoretically predicted.   We should finally note that for the same reasons as in the type I rumor spreading model, $S_k$, $I_k$ and $R_k$ may be easily recovered from the mean field solution.

Let us now describe why the dynamics of the recommendation spreading model are, at mean field level, SI epidemic dynamics. The answer can be seen by mapping  to a simpler problem, in the following way.   Define i.i.d. random variables $X_v$ for each $v\in V$, with $X_v \sim \mathrm{Bernoulli}((1+\lambda)^{-1})$, and remove from the graph $G$ all nodes $v$ with $X_v=0$.   The graph we are left with, which we call $G^\prime$, can be used to understand the $t=\infty$ state of a sample path for the recommendation model, in the following way:  $G^\prime$ consists of the possible nodes which will become As, if they have the chance to get infected.   Now, given a set of nodes which are A at $t=0$, we conclude that a final state for the dynamics of the recommendation spreading model is given by \begin{equation}
v(t=\infty) = \left\lbrace \begin{array}{ll}  \mathrm{A} &\ v \text{ not removed, in the same cluster as an initial A} \\ \mathrm{D} &\ v\text{ removed, connected to an A} \\ \mathrm{S} &\ \text{otherwise} \end{array}\right..
\end{equation}Furthermore, this final state has the same probability of occurring as the sum of all possible configurations of the ``removed node" model which lead to this same final state.   Given these states at $t\rightarrow\infty$, we can determine a sample path of the recommendation model by thus treating recommendation spreading as a SI epidemic on $G^\prime$ with spreading rate 1.   

This map to the SI epidemic on a reduced graph has a very interesting property, however -- it reveals that the recommendation spreading model actually has an ``epidemic threshold" in the following sense:   suppose that $G^\prime$ is almost surely a collection of clusters of O(1) nodes.   Then if, at $t=0$, an O(1) number of the nodes are A, at $t=\infty$ an O(1) number of nodes are A, implying that there is no recommendation ``epidemic."    A recommendation epidemic can only occur when the the cluster size grows with $N$.   This epidemic threshold does not occur within the context of mean field theory, and this is ultimately the crucial difference between the recommendation spreading model and the SIR-like models discussed above.

Given this understanding of the  late time dynamics, we now return to Figure \ref{zexp4}.  In particular (neglecting the constant factor making mean field theory differ from numerics for $\gamma=2.5$), we see that for very small $\lambda$, the only divergence from mean field theory is a finite size effect, because the probability that a giant cluster would not be present is presumably vanishingly small.  However, for larger $\lambda$, the probability that disconnected clusters occur becomes larger, and the value of $z$ at which the dynamics stops suggests the frequency with which such clusters occur.  For these larger values of $\lambda$, the dynamics of $z$ therefore deviates from mean field theory because the ending state of the dynamics is dependent on the existence and frequency of such clusters, and once the dynamics is dependent on graph structure, mean field theory breaks down.

\section{Conclusion}\label{concsec}

In this paper, we have shown that 4 simple models of irreversible dynamics on networks:  the SIR epidemic, the SI STD epidemic,  type I rumor spreading, and the new recommendation spreading model, have exact solutions at mean field level, and that these solutions hold up well in the appropriate regimes against numerical tests, differing at most by a constant scaling factor which is not too dramatic.\footnote{Why exactly such scaling factors occur is an open question -- part of the reason may be simplifications in the expression for $\theta(z)$, e.g.}   Thus, these results provide a far more thorough justification that mean field theory is a valid approximation scheme for these models than previous works.     Interestingly, proper regularization of divergences which can occur on heavy tailed degree distributions, such as those of scale free graphs, proved not only to be necessary mathematically, but to provide important physical insights as well.  

Ultimately, the SIR epidemic models, and the type IR or I rumor spreading models, are surely oversimplifications for realistic processes (and it is likely that realistic networks have far more structure than a simple ``mean field" scale free network), so the ultimate relevance of work such as this is to understand qualitatively why network structures can lead to  dramatic changes in the behavior of stochastic processes.  Towards this end, knowledge of an exact solution can help to solidify intuition that more heuristic approaches give, and can suggest phenomena that heuristic approaches may miss.  We have  showed that the exact solutions of mean field theory, which is often a valid approximation, provide all of the physical information of interest ($S_k$, $I_k$, and $R_k$) other than information dependent on the graph structure.   Finally, we were able to both provide theoretical explanations for many observed phenomena, as well as to postulate some new behaviors and observe them.

Recent work has mathematically proven some significant deviations from mean field behavior -- in particular, an absence of an epidemic threshold on scale free graphs of all degrees \cite{durrett2}.   While their results do not become relevant until $N\sim 10^{12}$, they showed that nonetheless mean field theory can sometimes be outright wrong, even on random graphs where physicists are most confident in mean field theory.    We hope that the (quite likely rare) existence of models whose mean field theory equations have exact solutions on arbitrary networks will provide key tests of when and where mean field theory is a valid approximation for simplified models of realistic networks and processes.   Future work should focus on understanding the extent to which our techniques may be applied to more complicated models, or other classes of models which may admit similar solutions, or focusing more in depth on some of the qualitative arguments we made (e.g., if $\tau_{\mathrm{spread}}\sim $ O(1)) which are not readily observable from our basic simulations.
\section*{Acknowledgements}\addcontentsline{toc}{section}{Acknowledgements}
I would like to thank Daniel Fisher, Greg ver Steeg and Jay Wacker for helpful comments and for encouraging me to continue past my initial calculations.

\bibliographystyle{plain}
\addcontentsline{toc}{section}{References}
\bibliography{epidemicbib}

\end{document}